# Focusing of Light by a Nano-Hole Array♣

*Fu Min Huang[1], Yifang Chen[2], F. Javier Garcia de Abajo[3] and Nikolay Zheludev[1]*

[1]EPSRC Nanophotonics Portfolio Centre, School of Physics and Astronomy, University of Southampton, SO17 1BJ, UK

[2]Central Microstructure facility, Rutherford Appleton Laboratory, Didcot, Oxon, OX11 OQX, UK

[3] Instituto de Optica, CSIC, Serrano 121, 28006 Madrid, Spain

**Abstract:** We demonstrate a conceptually new mechanism for sub-wavelength focusing at optical frequencies based upon the use of nano-hole quasi-periodic arrays in metal screens. Using coherent illumination at 660 nm and scanning near-field optical microscopy, ~290 nm "hot spots", were observed at a distance of ~12.5 μm from the array. Even smaller hot-spots of about 200 nm in waist were observed closer to the plane of the array.

The spatial resolution when imaging or patterning small objects with light is limited by diffraction to about half the wavelength. This is of primary importance in microelectronics and imaging as it translates directly into the size of the features that can be manufactured on a microchip or resolved by a microscope. Nano-holes and arrays of nano-holes in metal screens show a number of intriguing optical properties, most notably the extraordinarily high transmission of regular[1] and quasi-periodic[2] hole arrays and unusual polarization-sensitive effects.[3,4] Here we show for the first time that a quasi-crystal array of nano-holes in a metal screen can "focus" light at distances up to several tens of wavelengths from the screen. It produces bright foci of energy concentration, harvesting energy from a large number of holes in the array. The foci are sparsely distributed in the "focal plane". Using coherent laser illumination at a wavelength of 660 nm and scanning near-field optical microscopy (SNOM), “hot spots” with a FWHM of ~290 nm, which are separated from neighbouring foci by more than 5 μm, were observed at a distance of ~12.5 μm from the array. Even smaller hot-spots with a FWHM of ~200 nm were observed closer to the plane of the array.

The light focusing effect was observed with a quasi-crystal array of approximate 10-fold symmetry containing about 14,000 holes of 200 nm diameter. The overall diameter of the array was ~0.2 mm. The array comprised a Penrose-like pattern manufactured in accordance with the algorithm described in reference [5]. It was manufactured by electron beam lithography in a 100 nm aluminium film deposited on a silica substrate (see Fig. 1*i*). The high level of positional accuracy achieved in the manufacturing of the holes may be judged from a comparison of the far-field diffraction pattern produced by the array and the reciprocal lattice, calculated from the prescribed coordinates of the holes. Allowing for expected axial distortion and scaling these should be congruent mappings of each other, and indeed they are (compare Fig. 1*ii* and Fig. 1*iii*).

To map the field distributions created by the array we used a scanning near-field optical microscope (Omicron TwinSNOM R/T) operating in transmission mode, with light collected through a metal-coated tapered fiber tip with an aperture of about 120 nm. As optical sources we used either diode lasers operating at 660 nm and 635 nm or a continuum fiber laser and a selection of 40 nm bandpass filters. Fig. 2 shows characteristic field maps measured at different heights $h$ above the plane of the array for two different wavelengths $\lambda$, thus illustrating the variety of “photonic carpet” patterns generated by the quasi-crystal structure. In most cases, the carpets show elements of approximate 5-fold or 10-fold symmetry. In the immediate proximity of the sample (scans *i* and *vi*), the optical field concentrates at the holes. Further away from the array, the pattern changes dramatically and blurry, defocused patterns (*iii*, *iv*, *viii*, *x*) alternate with well-defined arrangements of hot-spots (*ii*, *v*, *vii*, *ix*) that appear at specific, irregular heights. At some heights and wavelengths bright foci, separated by a few microns from the neighbouring spots, are seen (scans *v* and *ix*). The patterns depend not only on the wavelength and height, but also on the distance between the focus of the illuminating lens and the quasi-crystal structure.

Fig. 3 shows a detailed scan of the foci at two characteristic distances, with the array illuminated at a wavelength of 660 nm. At a distance of about 5 μm from the array, a relatively dense pattern of hot-spots is seen. Here the bright foci are elliptical with a FWHM of ~375 nm along the polarization direction of the incident light and ~235 nm in the perpendicular direction. After deconvolution to account for the aperture size of the SNOM probe, the actual size of the hot-spots is estimated to be about 360 × 200 nm ($0.54\lambda \times 0.30\lambda$). A much clearer pattern of well-isolated hot-spots is seen at a distance of about 12.5 μm from the array. Here the scanned cross-sectional dimensions are 420 × 320 nm, and after deconvolution the spot size obtained is about 400 × 290 nm ($0.60\lambda\times0.44\lambda$). Thus, the structure produces bright foci of energy concentration by harvesting energy from a large number of holes. For example, in Fig. 2*v* the density of the brightest hot-spots (arranged in pentagons), which receive the majority of the energy, is some 40 times lower than that of the holes in the array.

This new phenomenon of "lensing" by the nano-hole array results from the partial reconstruction of the array's field in the diffraction zone, in a manner analogous to the classical Talbot effect[6] observed with periodic gratings, where diffraction leads to the reconstruction of the grating's field at periodic distances from the grating. In the Talbot effect a grating with period *a* images itself at multiples of the Talbot distance $h_T = a^2/\lambda$ when illuminated with a coherent plane wave. In the paraxial approximation, an infinitely long grating will be perfectly reconstructed at the Talbot distance. Montgomery[7] has shown that a wide range of patterns can image themselves, with linear periodic grating being only one example. A pattern will show full reconstruction at a distance *h* if its spatial frequencies are discrete and located in the reciprocal plane at rings of radii $\rho_m^2 = 1/\lambda^2 - (m/h)^2$, where *m* is an integer such that $0 \leq m \leq h/\lambda$. For instance, circular or linear diffraction gratings are self-imaging objects whose spectra in the reciprocal lattice are sets of equidistant rings or dots, respectively. The spectrum of the quasiperiodic array lies somewhere between the dotted spectrum of a linear grating and the rings of a circular grating. However, as the frequencies of the quasi-periodic array are located on circles (as shown in Fig. 1*ii*) the Montgomery condition is fulfilled for all spectral maxima, but not necessary simultaneously. Therefore the self-imaging distance $h$ will be different for different rings of maxima on the reciprocal lattice, and for different wavelengths $\lambda$. Moreover, when a regular grating is illuminated with a divergent beam, the reconstructed image is magnified, so the reconstructed field does not necessarily have the same period as the original grating. Similarly, when partial reconstruction of the quasi-periodic array takes place, the pattern of reconstructed hot-spots has the appearance of a scaled partial image of the array. Thus, reconstruction of a quasi-periodic array of holes is a complex process which may be envisaged as a superposition of a large number of partial reconstructions happening at different heights at the array. With varying height $h$, one will see a continuous evolution of partially reconstructed images of the array. What is important however, is that at some distances and wavelengths well defined sparsely distributed foci are seen.

To illustrate this process of quasi-periodic grating reconstruction, we have calculated the field intensity maps created by coherently oscillating dipoles oriented along the vector of the incident polarization $\boldsymbol{x}$ and located at the holes of the array with coordinates $\boldsymbol{R}_j$. The dipoles are assumed to be oscillating coherently as they are excited by a plane wave. In this case, the field on the other side of the array may be written as

$$\mathbf{E} = \sum_j \left[\nabla\partial_x + \hat{\mathbf{x}}\,k^2\right]\frac{e^{ik\left|\mathbf{r}-\mathbf{R}_j\right|}}{\left|\mathbf{r}-\mathbf{R}_j\right|}$$

where $\boldsymbol{r}$ is coordinate of the observation point and $k=2\pi/\lambda$. The results of these calculations are in a very good qualitative agreement with the experiment and revival that there are certain distances at which a pattern of focal Moreover, there are wavelengths and distances for which some of the bright spots have sub-wavelength diameters. For instance Fig.4 shows the formation of a hot spot of 220 nm in cross-section at the distance of 12μm from the sample. It is isolated in a 20μm×20μm region.

In conclusion, the focusing of coherent light by a quasi-crystal array of holes has been observed at distances ranging from a few microns to a few tens of microns from the plane of the array. With focused hot-spots of such high localization and brightness, imaging with subwavelength resolution can be achieved by scanning the object under investigation across the focal spot. Similarly, a single hot-spot appropriately isolated by a mask may be used as a "light pen" in high resolution photo-lithography.

**Acknowledgements.** This work was supported by the EPSRC and the Metamorphose NoE. We thank Nikitas Papasimakis & Alexander Schwaneke for taking far-field diffraction patterns of the array and Kevin Macdonald for help in preparing the manuscript.

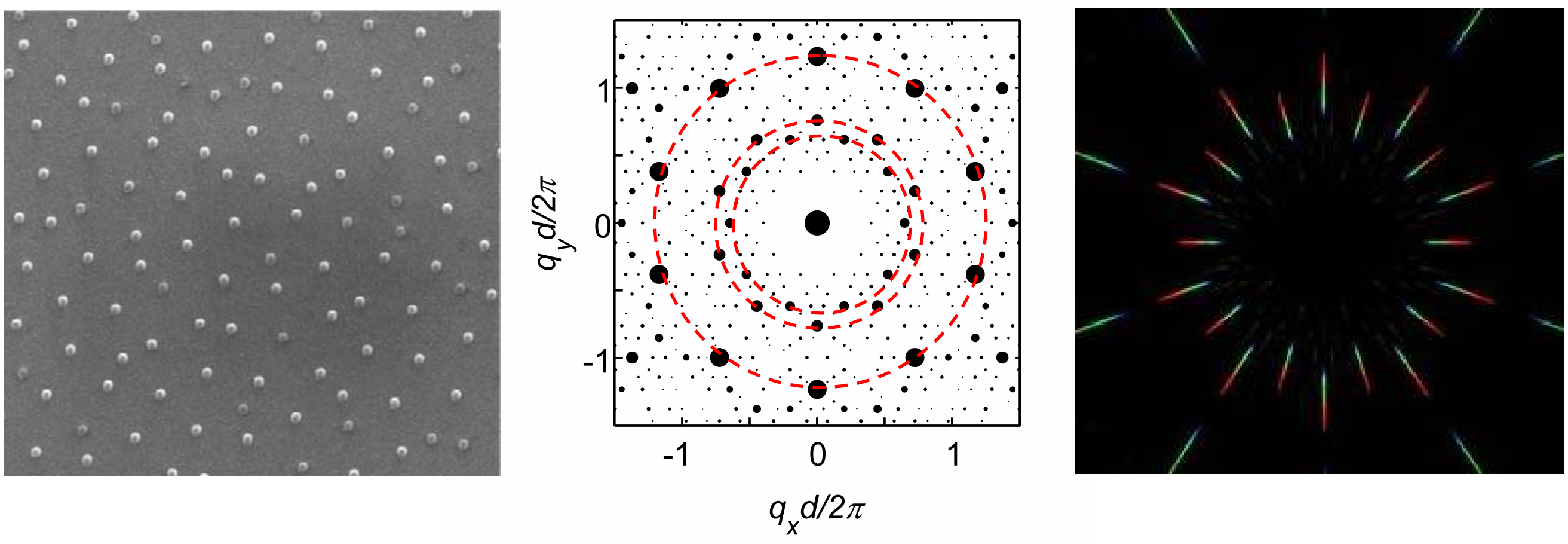

i ii iii

**Figure 1.** The quasi-crystal sample and its reciprocal lattice. (i) An SEM image of a fragment (20 × 20 μm) of the quasi-crystal array of holes; (ii) the reciprocal lattice of the quasi-crystal array wherein spot diameter is proportional to the magnitude of the spectral component. The minimum distance between two neighbouring holes *d* = 1.2 μm. Red dashed circles show several "partial" Montgomery rings. (iii) far-field diffraction pattern of the quasi-crystal array, obtained with white light illumination.

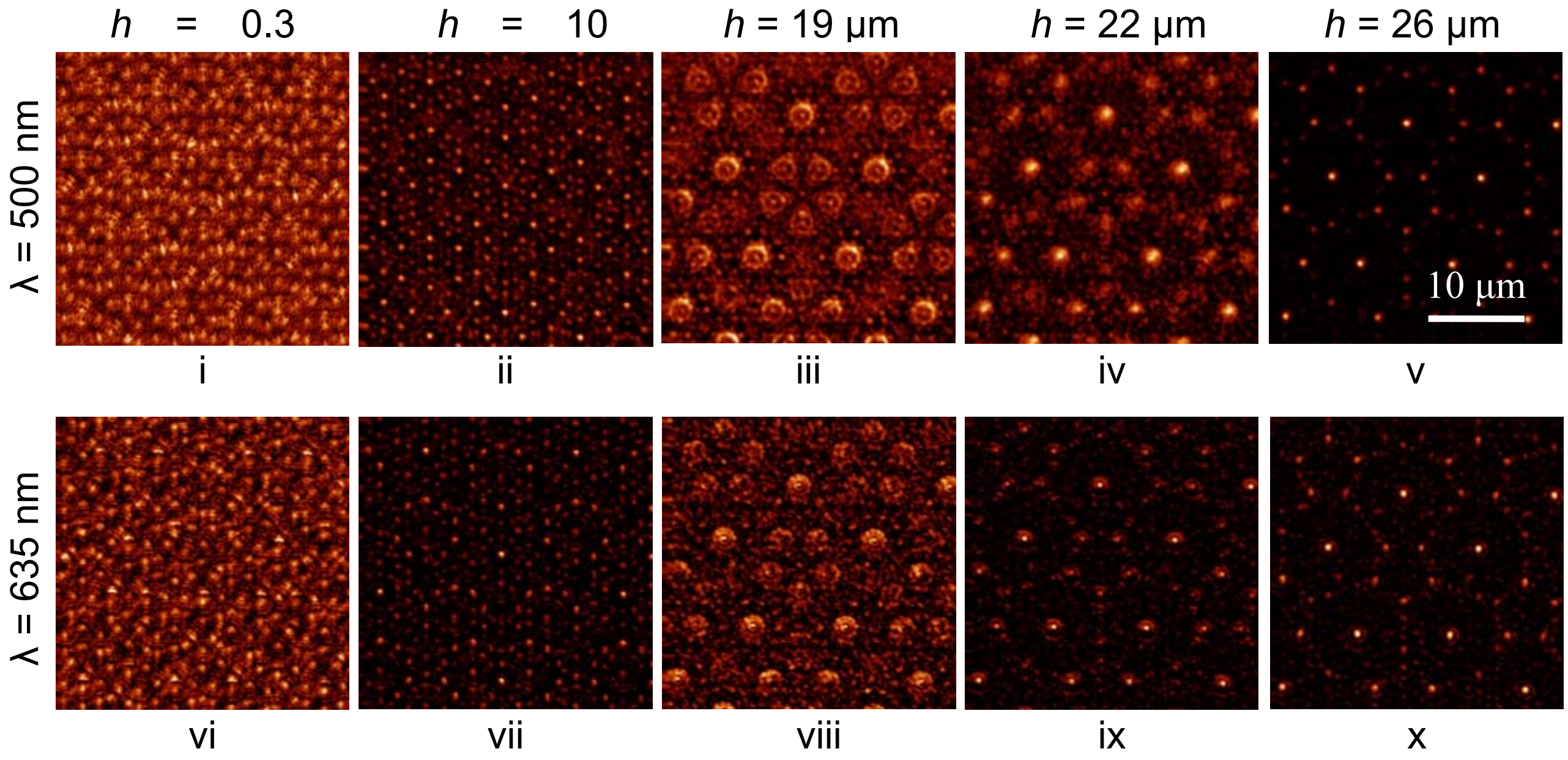

**Figure 2.** Field maps at different heights $h$ above the quasi-crystal array of holes. The images were obtained by SNOM using illumination wavelengths of 500 nm (upper row) and 635 nm (lower row). Note the highly isolated hot-spots observed at $h$ = 26 μm for the 500 nm excitation wavelength and at $h$ = 22 μm for the 635 nm excitation wavelength.

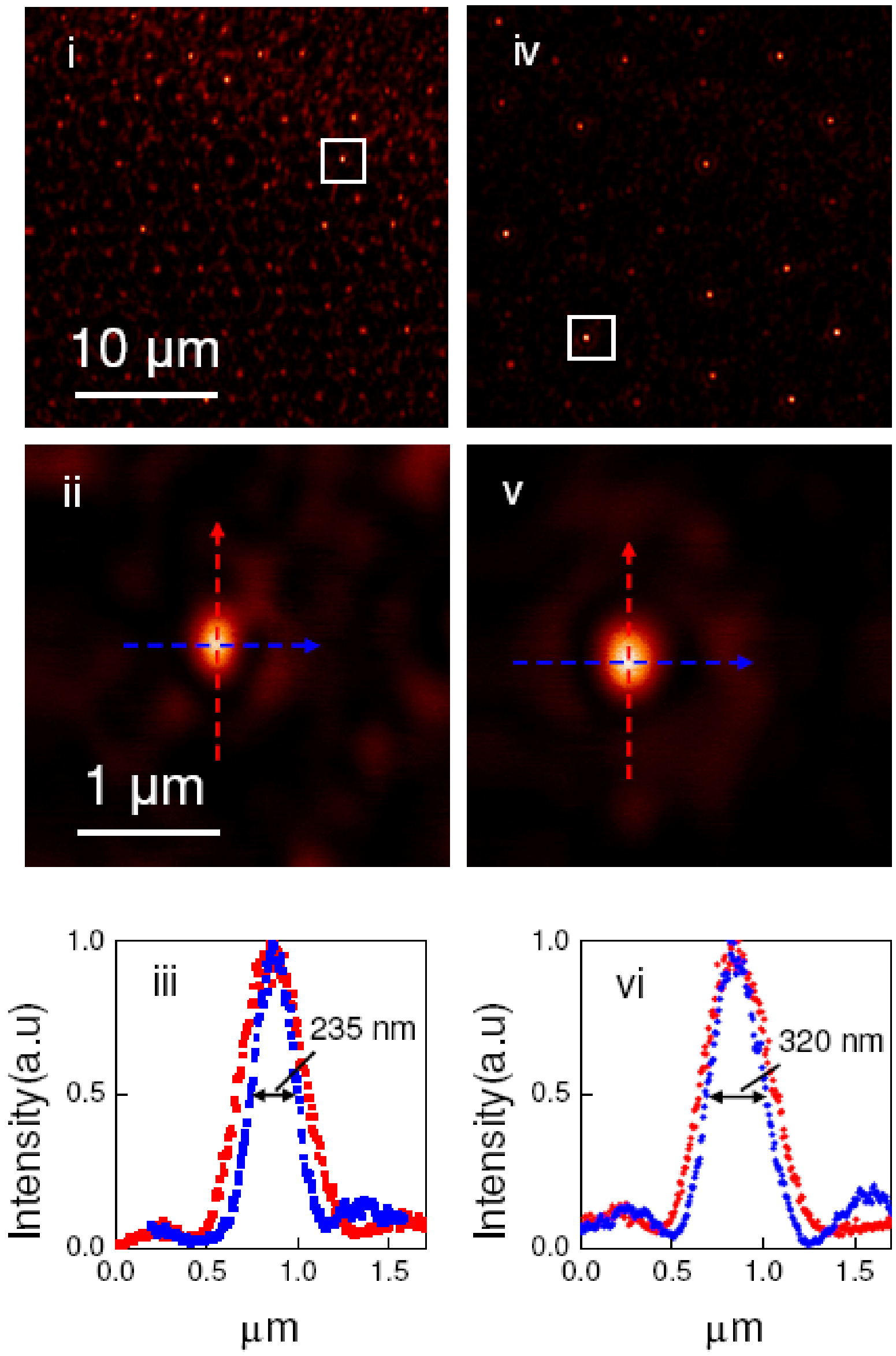

**Figure 3.** Subwavelength hot-spots formed at different heights above the quasi-crystal array. The illumination source is a diode laser with a wavelength of 660 nm. (i) field map at a height $h$ = 5 µm above the array; (ii), (iii) fine scan image of the hot-spot indicated in map (i) and corresponding profiles of the focus along directions parallel (blue dots) and perpendicular (red dots) to the polarization of the incident light; (iv) field map at a height $h$ = 12.5 µm above the array; (v), (vi) fine scan image of the hot-spot indicated in map (iv) and corresponding profiles of the focus along directions parallel (blue dots) and perpendicular (red dots) to the polarization of the incident light.

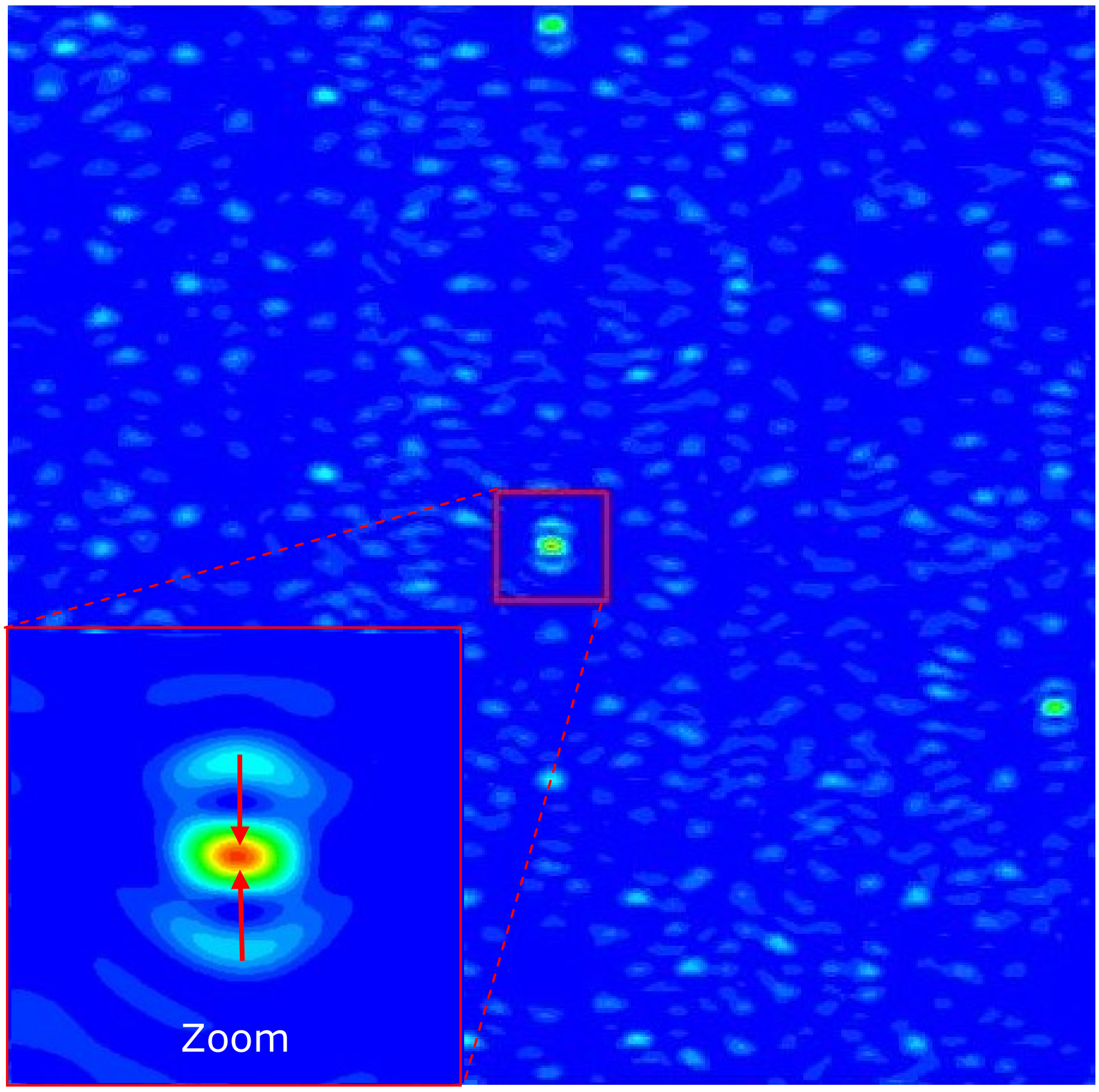

**Figure 4.** Theoretical calculations: subwavelength hot-spots formed at a height of 12μm from the sample above the quasi-crystal array for a plane wave excitation at a wavelength of 635 nm.